# Unexpected links reflect the noise in networks

**Authors:** Anatoly Yambartsev[1], Michael A. Perlin[2], Yevgeniy Kovchegov[3], Natalia Shulzhenko[4], Karina L. Mine[5], Xiaoxi Dong[2], Andrey Morgun[2*]

AY & AM equally contributed to this work.

*Correspondence to: anemorgun@hotmail.com; andriy.morgun@oregonstate.edu

**Affiliations:**

(1) Institute of Mathematics and Statistics, Department of Statistics, University of Sao Paulo, SP, Brazil.

(2) College of Pharmacy, Oregon State University, Corvallis, OR, United States

(3) Department of Mathematics, College of Science, Oregon State University, Corvallis, OR, United States

(4) College of Veterinary Medicine, Oregon State University, Corvallis, OR, United States

(5) Instituto de Imunogenética - Associação Fundo de Incentivo à Pesquisa (IGEN-AFIP), São Paulo SP, Brazil

**Abstract**

Gene covariation networks are commonly used to study biological processes. The inference of gene covariation networks from observational data can be challenging, especially considering the large number of players involved and the small number of biological replicates available for analysis. We propose a new statistical method for estimating the number of erroneous edges in reconstructed networks that strongly enhances commonly used inference approaches. This method is based on a special relationship between sign of correlation (positive/negative) and directionality (up/down) of gene regulation, and allows for the identification and removal of approximately half of all erroneous edges. Using the mathematical model of Bayesian networks and positive correlation inequalities we establish a mathematical foundation for our method. Analyzing existing biological datasets, we find a strong correlation between the results of our method and false discovery rate (FDR). Furthermore, simulation analysis demonstrates that our method provides a more accurate estimate of network error than FDR.

**Significance**

This study reports a discovery of a new property of interdependence between sign of correlation and direction of gene regulation for covariation networks first observed by us in cervical cancer. It appears to be universal as it has been further found in wide range of phenomena within biology and economics. Furthermore, the newly revealed property provides a basis for developing a method for measuring the proportion of erroneous edges in a network. This method stands out among standard approaches like the false discovery rate (FDR), because besides estimating an error it allows for the elimination of about half of all incorrect links in a network under a given statistical threshold.

**Introduction**

It is quite common, especially in biology, that in order to understand how systems transition from one state to another (e.g. from health to disease) scientists compare how parameters such as gene expressions, protein levels, or metabolite abundances differ between these states. One result of such a comparison is a list of parameters up- or down-regulated (due to the increase or decrease of some numerical value attributed to the parameter) from the first state to the second. In case of gene expression, these alterations represent a consequence of the two key factors: first, the original stimulus (e.g. mutation or environmental perturbation) that underlies the transition of a biological system from one state to another; and the second factor, a biological process that drives regulatory relations between individual genes independently on the presence of the stimulus. In other words, regulatory relations in biological systems (as well as many other systems) are not generally functions of the state but are rather pre-determined by biological roles of the components.

Most frequently, the components like genes are not regulated independently from each other; rather, they make up regulatory networks[1-5]. A common approach and the first step to the reconstruction of regulatory network structure is the inference of a correlation network built from parameters differentially abundant between two states. In particular, correlation (or, for the purposes of this paper, co-variation) networks are widely used in gene expression analysis. Co-variation network analysis works under the assumption that any edge (link) in a network, corresponding to a correlation between two parameters/nodes, is the empirical result of either direct or indirect (i.e. confounding) causal relationships, unless the edge is erroneously drawn (i.e. the observed correlation is an artifact of statistical error)[6-8]. Thus, we hypothesized that in a co-expression network there may be a relationship between the sign of correlation (i.e. positive or negative) of two regulated genes and the direction of their change between the two states (i.e. up or down-regulation). In this paper, we demonstrated the presence of this inter-dependence in different types of data, found that a departure from this relation reflects a proportion of erroneous edges in the regulatory networks, and developed a mathematical theory of this phenomenon.

**Results**

*The concept of unexpected correlations*

In order to verify whether there is a relationship between the direction of gene regulation and the sign of correlation we used a gene co-expression network from our recently published paper on network analysis in cervical cancer[9]. We felt that this network should provide excellent real data for this analysis, as it was constructed from a robust meta-analysis of five cancer gene expression datasets (GSE26342, GSE7410, GSE9750, GSE6791, GSE7803) and thus validated by large, independent sources. This network contained 738 nodes with 490 up and 248 down-regulated between cancer and normal tissues. These nodes were connected by 3161 edges with 2882 representing positive and 279 negative correlations. Relating these two types of information, we observed a strong association between the direction of gene expression change and the sign of correlation (Figure 1a). Positively correlated genes in ~98% cases had concordant increases or decreases in gene expression (up-up or down-down), and negatively correlated ones in ~92% of cases were regulated in opposite directions (up-down). At first glance we found surprising such a strong association and sought to further evaluate this phenomenon. Thus we focused on a part of this big network, which is a bi-partite network consisting of 626 correlations between gene-regulators and gene-targets[9]. In this smaller network in which correlation links could more obviously correspond to causal links (because gene-regulators have changed their expression as a result of chromosomal aberrations (Supp Fig. S1)), we found similar association between direction of correlation and gene regulation (Figure 1b).

We wondered whether such association can be generalized to other gene regulatory systems with two states (e.g. health and disease) and two types of regulation (stimulation and inhibition). In order to further investigate this, we propose a scheme in which we associate the sign of correlation (+/-) of each network edge with the direction (up/down) of gene regulation between system states. Sign association follows a simple set of rules:

- If there is a correlation between two "up" or "down" regulated genes (as in the top left panel in Figure 1c), the sign associated with the link is positive.
- If there is a link between an "up" regulated gene and a "down" regulated gene (as in the bottom left panel in Figure 1c), the sign associated with the link is negative.

The whole set of possible combinations of gene regulations and correlations are given in (Figure 1d). We hypothesize that correlations whose sign disagrees with the corresponding association are *erroneous*, i.e. they are the result of statistical error rather than causal relationships; or, they can be the results of an external/indirect influence, which is irrelevant for transitions between the biological system states. We will hereafter call such correlations unexpected (Figure 1d), and their proportion among all correlations in a network we abbreviate as PUC (the Proportion of Unexpected Correlations).

Since the original observation (Fig. 1) was made in complex system we also wanted to test the association between the sign of correlation and the direction of change in gene expression in the system where cause of gene regulation can be unambiguously defined.

For this, we employed a basic principle claiming that a result of experimental perturbation represents a *bona fide* causality relationship. In the same cervical cancer work, we had performed siRNA perturbation of gene LAMP3 (GSE29009), which was one of the key gene-drivers of the antiviral subnetwork. Our theoretical prediction would be that genes whose expression is affected by perturbation of the gene-driver (i.e. LAMP3) *in vitro* and correlated to the expression of the gene-driver in the original cancer data should present correlations of the expected sign. For example, if a gene was down-regulated by LAMP3 siRNA, it is expected to be positively correlating to LAMP3 in the cancer gene expression data and vice versa (i.e. if gene is up after siRNA treatment correlation should be negative). Thus we analyzed if the direction of regulation of genes affected by LAMP3 siRNA in the cell line was corresponding to the sign of correlation between each gene and LAMP3 in four cervical cancer datasets (GSE7410, GSE9750, GSE6791, GSE7803). In these datasets, we observed that almost all correlations between LAMP3 and genes whose expression was affected by LAMP3 siRNA had correlation signs concordant to the directions of gene regulations due to siRNA treatment (Figure 1e). Thus, this data provides the additional experimental support for our hypothesis about non-random interdependence between sign of correlation and direction of gene regulation.

*Mathematical formalism relating causation and the sign of correlation*

Encouraged by these results, to better understand the properties of this new metric (PUC) we went further to establish a mathematical framework for its application. Although concept of PUC can be formulated and tested empirically without mathematical theory, a rigorous formalization of PUC is necessary to provide a theoretical basis for its application and to establish potential limits for its generalization if there is any. Our hypothesis that unexpected correlations are erroneous can be rigorously proven for systems that transition between two stable states with two types of regulations between parameters: stimulation and inhibition. Herein, as an illustration we provide a proof of our hypothesis in the bounds of a simple mathematical model, namely that of Bayesian networks[9] with two equilibrium states and linear dependences between nodes (the proof for more general case is provided in Section II.2 of the Supporting Material). In order to formulate our results we need to introduce some mathematical notation.

Consider a regulatory network without loops (i.e. a directed acyclic graph, DAG) represented by a graph $G = (V, E)$. Any edge $e \in E$ is an ordered pair of vertices (nodes) $e = (v, w) \in V^2$. The order of an edge represents the direction of causality in a regulatory network (that is, an order $(v, w)$ implies that $v$ regulates $w$). For any node $v$ we associate the set of its parents as $pa(v) := \{u \in V: (u, v) \in E\}$. We define the set of root-nodes $gf(G)$ for the graph $G$ as the set of all nodes without parents: $gf(G) := \{v \in V: pa(v) = \emptyset\}$.

Let graph $G$ be weighted, meaning that every edge $e = (v, w) \in E$ has an associated label (weight) $c_{vw} \in R$. With every node $v \in V$ we associate a random variable $M_v$. The distribution of random variables is given by their respective structural linear equations:

$$M_v = \sum_{w \in pa(v)} c_{wv} M_w + \varepsilon_v$$

Here $\varepsilon_v$ are mutually independent and identically distributed with mean 0 and variance $\sigma^2$. We suppose homoscedasticity ($Var(\varepsilon_v) = \sigma^2$, variance is same for all vertices $v$) for simplicity, to make the proof of Lemma 1 more clean and clear. The proof is also straightforward if we allow heteroscedasticity but with uniformly bound variances: $\exists \sigma^2 : Var(\varepsilon_v) \leq \sigma^2 \forall v \in V$.

In the previously discussed biological framework, a graph $G$ represents the entire gene expression network. A node $v$ represents some gene, which has an expression level $M_v$. An edge $e = (v, w)$ represents a causal link between two genes $v$ and $w$ in which the expression of $w$ is regulated by $v$. The sign of $c_{vw}$ reflects the direction of regulation: a negative (positive) sign corresponds to inhibition (stimulation). The parents of $v$ are simply all genes which regulate $v$ and the root-nodes of $G$ are the primary regulators of the entire network, i.e. the genes at the top of the regulatory chain.

For simplicity, we consider a regulatory network with only one root-node ($|gf(G)| = 1$), denoted by the vertex $o$. The case with more than one root-node is covered by the general model considered in Supporting Material, see Section II.2. Let $M_v^{(P)}$ and $M_v^{(Q)}$ denote the expressions of node $v$ in two distinct equilibrium states $P$ and $Q$. We will use the notation $X_v^{(S)}$ to denote a variable $X$ associated with node $v \in V$ in the state $S \in \{P, Q\}$. For any $v$ we denote the changes in expression between states as $\Delta_v = EM_v^{(P)} - EM_v^{(Q)}$, where $E$ denotes the expectation value (mean) of corresponding variable.

The mathematical definition of expected and unexpected links, given heuristically in the introduction, is formally expressed in the following definition.

*Definition. An edge $e \in E$ is called an expected link between nodes $v, w \in V$ if and only if $\Delta_v \Delta_w cov\left(M_v^{(P)}, M_w^{(P)}\right) > 0$ and $\Delta_v \Delta_w cov\left(M_v^{(Q)}, M_w^{(Q)}\right) > 0$. Any edge which is not an expected link constitutes an unexpected link.*

This definition effectively states that the directions of regulation of two genes between two states should agree with the sign of the correlation between them within each state.

First we proved the lemma that states that unexpected signs of correlations are result of noise (the proof is given in Section II.1 of the Supporting Material).

*Lemma 1. For any finite DAG with linear structural equations there exists some $\sigma_0$ such that if $Var(\varepsilon_v) < \sigma_0^2$ for all $v \in V$ then there are no unexpected links.*

Another very important property of the concept of unexpected links is that PUC represents and identifies approximately half of all erroneous correlations:

$$2E(PUC) \approx E(\text{total proportion of false positive links}).$$

A formal proof of this statement is given in Section III.3 of the Supporting Material, as well as an explanation for why this makes intuitive sense. The basic idea lies in the observation that false edges are, in principle, equally likely to satisfy the conditions for expected correlation as they are to satisfy the conditions for unexpected correlations.

*Unexpected correlations reflect the noise in real and simulated networks.*

Our mathematical analysis proved that in regulatory networks unexpected correlations must have appeared as a result of noise within the network and that the proportion of unexpected correlation thus reflects the noise level in a network.

Mathematical models are restricted by the domain of their assumptions, which limits their applicability. Thus, although we have empirically observed a small PUC in a high confidence cervical cancer network (Figure 1a-b), we wanted to verify whether this correspondence would still hold in different settings. We therefore analyzed 24 additional data sets retrieved from the BRB Array Tools Archive (see supporting material) providing gene expression network transitions in different types of cancer, and found that PUC strongly correlated with FDR (correlation coefficient of 0.87 CI95% 0.8117-0.9416). Turning our attention to phenomena other than cancer, we also analyzed the gene expression network perturbed as a result of colonization of intestinal tissue with normal microbiota (i.e. the mix of microorganisms that live in the gut). In these data (GSE60568)[10] and again found that PUC is highly correlated with FDR (Figure 2e).

Thus, the observation of strong correlation between FDR and PUC in multiple datasets from diverse biological settings in two different species (Homo sapiens and Mus musculus) provides additional support for our prediction that PUC, similarly to FDR, quantitatively reflects network error.

An important question, however, is whether PUC brings any advantage over the standard approach to measuring the proportion of erroneous edges in a reconstructed regulation network (i.e. FDR). Real data makes such a comparison difficult, because though both methods of analysis will return values for network error, there is not necessarily any obvious way to determine which is more accurate; i.e. in real data, the actual regulatory network is not known.

To investigate the behavior of PUC in a "controlled environment" we simulated networks using two approaches. We have used simple Bayesian networks[11] and GeneNetWeaver (networks simulated using stochastic and/or ordinary differential equations)[12] generated networks as models for gene regulation. In order to compare the effectiveness of PUC and FDR, two regulatory networks are constructed and simulated independently, and both networks' node expression levels combined into one data set. In a correlation network constructed from the simulated data, any correlations (link) between nodes from independent networks are known to be erroneous (Figure 2a). This design allows for a true measure of network error against which to compare PUC and FDR analysis results.

In order to determine which method (FDR or PUC) better quantifies error, we look at all three measures of error (FDR, PUC, and the true error) and compare the accuracies of FDR and PUC. The results of both types of simulations suggest that PUC is more accurate than FDR in estimating true error, although there is a strong correlation between the two metrics (Figures 2b,c).

The FDR family of methods is the most popular procedure for large-scale p-value correction for multiple hypotheses[13-17]. All these FDR methods, however, ignore the dependence structure between hypotheses, which leads to the fact that FDR is an overly conservative approach (i.e. it overestimates the number of false positives).

In the case of regulatory networks, each edge constitutes a hypothesis; interdependency of regulatory network hypotheses manifests in indirect regulation between genes. Indeed, this is exactly the case with co-variation networks, in which it is possible to find numerous indirect pathways with only a few direct links.

Using PUC as a measure of error, however, does not require any assumption of hypothesis independence. PUC may thus be more accurate than FDR for error estimation in co-variation networks with a large number of interconnected nodes. The "degree of dependency" between hypotheses also depends on the size and number of sub-networks that compose a network. A network made up of twenty sub-networks consisting of twenty nodes each should have a lower degree of hypothesis interdependency than a single network consisting of four hundred nodes lacking any well-defined sub-networks.

In order to catch this effect we simulated various networks with 400 nodes in disjoint sub-networks, each with an equal number of nodes (for example, 20 disjoint sub-networks with 20 nodes each). While both types of simulations (Bayesian and GeneNetWeaver networks) showed overall more accurate results for PUC, in Bayesian networks we also observed lower efficiency of FDR for large networks (Figure 2d). This effect, however, was not pronounced in GeneNetWeaver- simulated networks (Figure S4). Furthermore, we obtained similar results (Figure 2e) by using another version of FDR which was designed to correct for hypothesis interdependency – Benjamini-Yekutieli, FDR-BY[18].

*PUC in a non-biological system*

The fact that we could mathematically prove the relationship between unexpected correlations and network error suggests that this principle could be widespread beyond gene interactions in biological systems. As a proof-of-concept of PUC's generality, we turned our attention to economics. The justification for this choice of subject relates to the presumption that economic systems, similarly to biological systems, are governed by cause-effect relationships and can, by extension, be described by regulatory networks. We analyzed 1503 parameters retrieved from World Bank economic databases for the year 2008 in 193 countries in such areas as business, education, health, etc. (details provided in the Supplementary Information). Parameters with bimodal distributions defined distinct states of economic networks for any given country. Figure 2f shows PUC for parameter correlation networks with different FDR thresholds in which a particular parameter (expenditure per student on primary education as a percent of GDP per capita) defined distinct network states; Figure S2 (in the Supplementary Information) shows similar graphs for various other parameters. As expected, these networks demonstrated a high concordance between the network errors given by PUC and FDR. This result supports the idea that the concept of unexpected correlations can be extended to a large variety of causal networks and that measurement of the proportion of unexpected correlations (PUC) can improve network analysis in a variety of scientific disciplines.

*Estimating error using PUC*

The entire procedure of PUC for calculating network error is as such: first, all correlations in a differential expression list are ranked by p-value. A network is constructed with edges consisting of correlations within an arbitrary p-value threshold (e.g. 0.01). Unexpected links are identified, counted, and removed from the network. The final measure of error in the remaining network is given by, where  and  are respectively the numbers of total and unexpected links in the network prior to removal of unexpected links. The reason for this formula is explained in the last paragraph of mathematical formalism section, and has to do with the fact that the number of unexpected links in a network is approximately equal to half of the total number of false links.

**Discussion**

The growth of molecular biology has advanced such that we can measure the expression of thousands of genes simultaneously. Simply measuring the expression of multiple individual genes, however, is insufficient to describe a systems issue such as complex diseases. To relate gene expression to physiological states (e.g. disease) and other variables in an organism's environment we utilize gene expression networks. These networks enable more intelligent identification of molecular subtypes of diseases and molecular targets for treatment. The reconstruction of gene expression networks, however, is not easily accomplished. Constructing reliable gene expression networks with current methods requires obtaining large data sets because a large number of hypotheses are required to be tested for network inference.

Although the False Discovery Rate (FDR - Benjamini-Hochberg) is the most popular multiple hypothesis correction method, its application for network inference is a conservative procedure and makes the often unfitting assumption of the independence between correlations in gene networks. There are less popular versions of FDR, for example Benjamini-Yekutieli[18], which take into account various dependence structures between the hypotheses under consideration, but the usage of this did not demonstrate any significant advantage over PUC (see Figure 2e). Consequently, these corrections tend to have a high rate of false negative discovery (i.e. low power) and require vast sample sizes in order attain desirable degrees of certainty about reconstructed networks. There is thus a critical need for more powerful methods of estimation of false positive connections between genes in co-expression networks.

In this study we have revealed and mathematically proved a new feature of co-expression networks. This feature is based on the natural notion that any correlation has direct or indirect causal components and noise components. In the case when causal components prevail over noise, the sign of a correlation between two genes should be related to their up- or down- regulation of the genes between two states (Figure 1).We first observed this relation empirically in gene expression datasets[9,19], and subsequently in macroeconomic data (see Figure 2f and Figure S2). The observation of this network feature (relation between sign of correlation and direction of change) in data of such a different nature (biology and economics) suggests that this relation is a universal property of covariation networks.

We proposed using this relation for identifying false connections in co-variation networks, increasing network accuracy, an estimating total network error. This approach demonstrates clear advantage over the classic method (FDR) not only by providing better estimates of error in large covariation networks, but also by allowing the removal of approximately half of all erroneous edges. The fact that PUC demonstrates similar behavior to standard methods of analysis (i.e. PUC has a strong correlation with FDR) in both real and simulated networks further supports the use of this adopted modeling approach. Indeed, certain questions can only be answered using a modeled system. We had to use simulated networks where we know the exact number of false links to compare FDR and PUC.

The identification of unexpected correlations has two primary impacts. Firstly, it provides a new method to estimate the proportion of erroneous links in a network. Secondly, it allows for the *removal* of approximately half of the erroneous edges in the network (namely, those that are unexpected), decreasing their proportion by a factor of two and thereby improving the overall accuracy of the reconstructed network. The final value of network error consists of an estimated proportion of remaining false positive correlations.

The concept of expected and unexpected correlations that we introduced is closely related to the concept of monotone causal effects and the covariance between them[20]. The rules we proved for linear relations should therefore hold for any monotone relationships; this idea is expanded in Section II.2. of the Supporting Material, and the framework of PUC extended to a broader class of networks than those mentioned thus far.

We must also address how non-monotonicity affects the notion and application of unexpected correlations. The concept of non-monotonicity can be exemplified for our problem as different types of relationships in two network states, such as a negative correlation between parameters in one biological state and a positive correlation in another. In such cases, despite violation of monotonicity, we expect unexpected correlations to arise primarily due to noise, rather than the change in relationships. Nonetheless, we demonstrated (see Section II.4. of the Supporting Material) that there is no evidence for non-monotonicity to suggest that these exceptionally rare non-erroneous correlations are in fact responsible for the observed changes in gene expression between states of a biological system. Therefore, because the ultimate goal of network inference is actually to model and understand the transition of biological system from one state to another, we can safely remove these unexpected correlations from the reconstructed network for independent reasons (i.e. that they do not have causal contribution to system state transition).

We believe that this work, besides revealing a new feature of covariation networks, introduces an entirely new way of dealing with error in their reconstruction. Indeed, statistical methods employed for such problems normally estimate an error, but cannot detect erroneous edges. We propose a method that besides (according to simulations, potentially superior) error estimation allows for identification and removal of approximately half of total network error. Thus, the identification and removal of unexpected correlations decreases the proportion of irrelevant and erroneous connections and strongly increases the power of network inferences.

**Acknowledgements:** We thank Eric Zubriski for developing a code and running some simulations, Chris Sullivan, from Oregon State University for help in setting-up computation infrastructure at CGRB (Center for Genomic Research and Biocomputing). We also thank Amiran Dzutsev, Steve Ramsey, Lina Thomas and Jesse Zaneveld for critical reading of the manuscript. AY was supported by FAPESP (grant 2012/06564-0); YK and AM are supported by NSF grant - Award Number: 1412557.
**Contributions:** AY conceived original idea , developed mathematical model, supervised simulation analyses, and drafted the manuscript; MAP analyzed economical and cancer data, performed simulations analysis with GeneNetWeaver and drafted the manuscript; YK proved generalized mathematical model, drafted corresponding part of manuscript; NS provided data for microbiota colonization, participated in analyses of cancer data and drafted manuscript; KLM participated in analysis of cancer data; XD analyzed cancer and gene silencing data; AM conceived original idea, analyzed data, supervised and led the whole project, drafted the manuscript.

We declare that the authors have no competing interests as defined by Nature Publishing Group, or other interests that might be perceived to influence the results and/or discussion reported in this paper.

**Figure 1:** Sign of correlations corresponds to the direction of change in regulatory networks. a) Percentage of positive and negative correlations for pairs of up-regulated (up) and down-regulated (down) genes observed in the network from Mine et al., 2013; b) number of positive and negative correlations between pairs of target and regulator genes in relation to their up- or down- regulation in cervical cancer data; c) examples of regulatory (left panels) and erroneous (right panels) connections between genes X and Y; d) possible combinations of gene regulations and correlations with the interpretation of connection; e) percentage of expected and unexpected connections between LAMP3 and other differential expressed genes in cervical cancer corresponding to genes regulated after knockdown of LAMP3 in four datasets: Beiwenga (GSE7410), Pyeon (GSE6791), Zhai (GSE7803), Scotto (GDS3233).

**Figure 2:** Comparison of PUC and FDR. a) Two regulatory networks are simulated independently, then both networks' node expression levels combined into one data set. In a correlation network constructed from the simulated data, any correlations (links) between nodes from independent networks are known to be erroneous; Bayesian simulations (b), as well as gene regulatory simulations performed GeneNetWeaver (c) suggest that PUC more accurately reflects network error than FDR (Benjamini-Hochberg, FDR-BH); as network size grows, PUC more accurately reflects network error than FDR-BH (d) or its variation with multiple hypothesis under dependence called FDR Benjamini-

Yekutieli (FDR-BY) (e); PUC correlates with FDR in both gene expression (f) and macroeconomic (g) data.

# References

1. Butte, A. J., Tamayo, P., Slonim, D., Golub, T. R. & Kohane, I. S. Discovering functional relationships between RNA expression and chemotherapeutic susceptibility using relevance networks. *Proceedings of the National Academy of Sciences of the United States of America* **97**, 12182-12186, doi:10.1073/pnas.220392197 (2000).
2. Opgen-Rhein, R. & Strimmer, K. From correlation to causation networks: a simple approximate learning algorithm and its application to high-dimensional plant gene expression data. *BMC systems biology* **1**, 37, doi:10.1186/1752-0509-1-37 (2007).
3. Pe'er, D. & Hacohen, N. Principles and strategies for developing network models in cancer. *Cell* **144**, 864-873, doi:10.1016/j.cell.2011.03.001 (2011).
4. Shulzhenko, N. *et al.* Crosstalk between B lymphocytes, microbiota and the intestinal epithelium governs immunity versus metabolism in the gut. *Nature medicine* **17**, 1585-1593, doi:10.1038/nm.2505 (2011).
5. Margolin, A. A. *et al.* ARACNE: an algorithm for the reconstruction of gene regulatory networks in a mammalian cellular context. *BMC bioinformatics* **7**, S7 (2006).
6. Pearl, J. An introduction to causal inference. *The international journal of biostatistics* **6**, Article 7, doi:10.2202/1557-4679.1203 (2010).
7. Pearl, J. *Causality: models, reasoning and inference*. Vol. 29 (Cambridge Univ Press, 2000).
8. Reichenbach, H. *The direction of time*. Vol. 65 (Univ of California Press, 1991).
9. Mine, K. L. *et al.* Gene network reconstruction reveals cell cycle and antiviral genes as major drivers of cervical cancer. *Nature communications* **4**, 1806, doi:10.1038/ncomms2693 (2013).
10. Morgun, A. *et al.* Uncovering effects of antibiotics on the host and microbiota using transkingdom gene networks. *Gut*, doi:10.1136/gutjnl-2014-308820 (2015).
11. Thomas LD, F. V., Yambartsev A. Building complex networks through classical and Bayesian statistics-A comparison. *arXiv:1409.2833* **1490,**, 323 (2012).
12. Schaffter, T., Marbach, D. & Floreano, D. GeneNetWeaver: in silico benchmark generation and performance profiling of network inference methods. *Bioinformatics* **27**, 2263-2270, doi:10.1093/bioinformatics/btr373 (2011).
13. Benjamini, Y., Hochberg, Y. Controlling the false discovery rate: a practical and powerful approach to multiple testing. *Journal of the Royal Statistical Society. Series B* **57** (1995).
14. Genovese, C. & Wasserman, L. Operating characteristics and extensions of the false discovery rate procedure. *Journal of the Royal Statistical Society: Series B (Statistical Methodology)* **64**, 499-517 (2002).
15. Efron, B., Tibshirani, R., Storey, J. D. & Tusher, V. Empirical Bayes analysis of a microarray experiment. *Journal of the American statistical association* **96**, 1151-1160 (2001).
16. Storey, J. D. A direct approach to false discovery rates. *Journal of the Royal Statistical Society: Series B (Statistical Methodology)* **64**, 479-498 (2002).
17. Storey, J. D. The positive false discovery rate: A Bayesian interpretation and the q-value. *Annals of statistics*, 2013-2035 (2003).
18. Benjamini, Y. & Yekutieli, D. The control of the false discovery rate in multiple testing under dependency. *Annals of statistics*, 1165-1188 (2001).
19. Skinner, J. *et al.* Construct and Compare Gene Coexpression Networks with DAPfinder and DAPview. *BMC bioinformatics* **12**, 286, doi:10.1186/1471-2105-12-286 (2011).
20. VanderWeele, T. J. & Robins, J. M. Signed directed acyclic graphs for causal inference. *Journal of the Royal Statistical Society: Series B (Statistical Methodology)* **72**, 111-127 (2010).

# Supplementary Information:

# Unexpected links reflect the noise in networks

**Authors:** Anatoly Yambartsev[1], Michael A. Perlin[2], Yevgeniy Kovchegov[3], Natalia Shulzhenko[4], Karina L. Mine[5], Xiaoxi Dong[2], Andrey Morgun[2].

## I. Experimental procedures

### I.1. Statistically significant correlations between differentially expressed genes (DEGs) and show expected signs

In our recent study (Nature Commun. 2013;4:1806) we have shown that key drivers of cervical carcinogenesis are located in regions of frequent chromosomal aberrations and that these genes cause most of the alteration in gene expression in cervical cancer. Therefore, in order to evaluate whether statistically significant correlations between DEGs which result from known causal relations follow our prediction we performed the following analysis:

First, we selected two groups of genes from DEGs discovered in our previous study: 1) genes in which it has been determined that chromosomal aberrations are responsible for the change in regulation; and 2) genes located in regions in which aberrations are rare, defined by FqG – FqL between -0.1 and 0.1 (Figure S1). Next, we analyzed gene co-expression in tumors samples in order to find correlations between those two groups of DEGs. We found 626 correlated gene-gene pairs with FDR 5%. We used data from the following datasets for our meta-analysis (performed as described in Nature Commun. 2013;4:1806): GSE26342, GSE7410, GSE9750, GSE6791, GSE7803. In brief, we calculated correlations within the tumor samples of each dataset. If correlations presented the same sign in all datasets, then we calculated a corresponding Fisher meta-analysis p-value. We then computed the FDR for these correlations. The results provided support to our hypothesis that significant correlations should have "expected" signs. Indeed, 95% (594 of 626 total pairs) of significant correlations had expected signs.

### I.2. PUC correlates with FDR in macroeconomic data

The macroeconomic data we analyzed was a combination of all data for the year 2008 on official UN member states in the following World Bank databases: Doing Business, Education Statistics, Gender Statistics, Health and Nutrition Population Statistics, IDA Results Measurement System; Poverty and Inequality Database, World Development Indicators and Global Development Finance. From this data set, we removed all duplications of macroeconomic parameters, as well as all parameters for which data only existed for $\leq 25$ countries. Of the remaining parameters, we used a dip test to determine those which were non-unimodal with a p-value of $< 2.2\times10^{-16}$. From the resulting set of parameters, we selected several with bimodal distributions, each of which we used to define two distinct states of a macroeconomic parameter network. We then computed PUC for parameter correlation networks at different FDR thresholds using each of these definitions. The results of these calculations are shown in Figure S2.

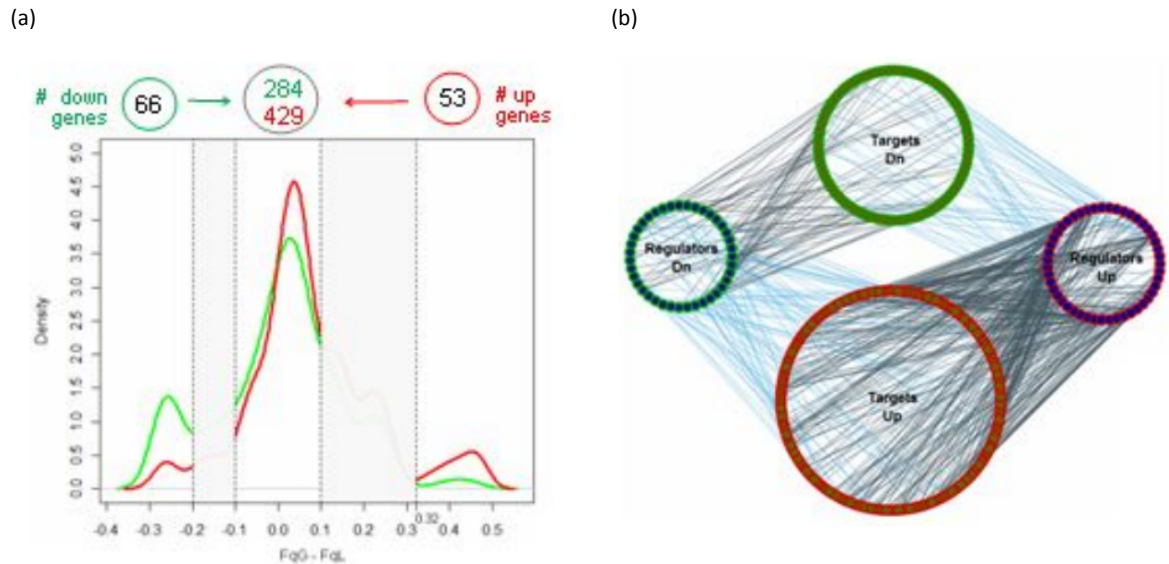

**Figure S1: Genes directly regulated by chromosomal aberrations can also in turn regulate genes located outside of the aberrations.** (a) Genes regulated by chromosomal aberrations in the expected direction (located in the regions $FqG - FqL < -0.2$ or $FqG - FqL > 0.3$) were considered as potential regulators, and genes located within the regions of very rare aberrations ($|FqG - FqL| \leq 0.1$) were considered to be potential targets. The green (red) line represents up-regulated (down-regulated) genes. (b) The reconstructed regulatory network with correlations in agreement with gene expression. The two green (red/purple) circles are made of up down-regulated (up-regulated) nodes, the middle (side) circles are made up of targets (regulators), and the black (cyan) lines represent positive (negative) correlations.

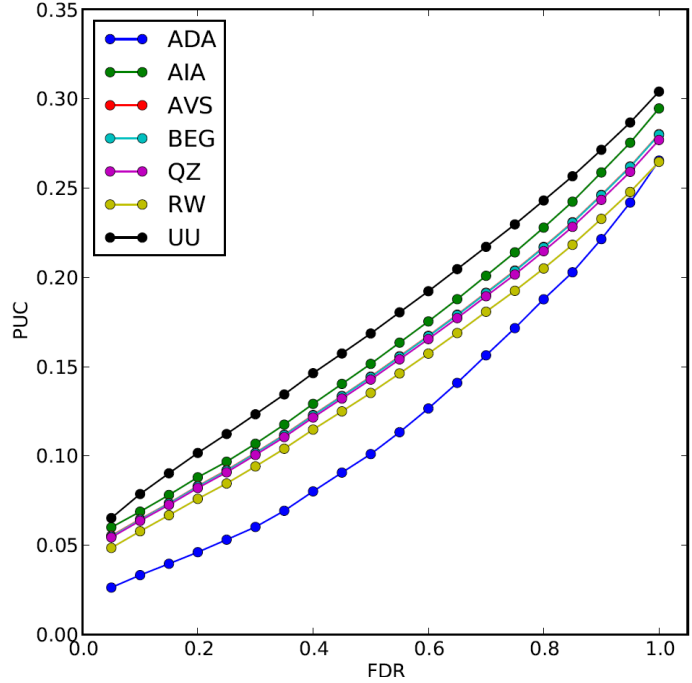

**Figure S2: PUC and FDR correlate strongly when reconstructing macroeconomic networks using various bimodal parameters to define system states.** Parameters shown: ADA - Duration of compulsory education; AIA - Cause of death, by communicable diseases and maternal, prenatal and nutrition conditions (% of total); AVS - Manufactures exports (% of merchandise exports); BEG - Educational expenditure in pre-primary as % of total educational expenditure; QZ - Private credit bureau coverage (% of adults); RW - Strength of legal rights index; UU Passenger cars (per 1,000 people)

## II. Theoretical basis.

Here we provide some formal definitions of concepts used in the paper and all necessary proofs. This section consists of four parts: 1) we introduce the mathematical machinery for PUC using Bayesian networks; 2) we generalize the previous formalism to handle a broader set of cases; 3) we demonstrate that PUC reflects half of total network error; and 4) we address concerns with network non-monotonicity.

### II.1. PUC on Bayesian networks.

In order to apply the new concept of noise estimator we use Bayesian Networks as a convenient model for gene expression. Let $G = (V, E)$ be some network, which is directed acyclic graph (DAG). Any edge $e \in E$ is an ordered pair of vertices $e = (v, w)$: and direction of edge is from the first vertex $v$ to the second vertex $w$. We assume that the graph is weighted graph – any edge $e = (v, w)$ has its labels (weight), $c_{vw}$, which is some real number $c_{vw} \in \mathbb{R}$. For any node $v$ we associate the set of parents of the node $v$:

$$pa(v) \coloneqq \{w \in V : (w, v) \in E\} \tag{1}$$

We define the set of grandfathers for the graph $G$:

$$gf(G) := \{v \in V : pa(v) = \emptyset\} \quad (2)$$

With any node (gene) $v \in V$ we associate the random variable (gene expression) $M_v$. The random variables satisfy the following linear relations (structure equations): for any $v \notin gf(G)$

$$M_v = \sum_{w \in pa(v)} c_{wv} M_w + \varepsilon_v, \quad (3)$$

where $\varepsilon_v$ are i.i.d. random variable (intrinsic noise) with mean 0 and variance $\sigma^2$. Moreover, for simplicity we suppose that there exists only one grandfather $|gf(G)| = 1$ and let us denote it as a vertex $o$.

A path $\pi(v, w)$ of length $n$ from a vertex $v$ to a vertex $w$ is a sequence of edges $e_i = (v_i, v_{i+1}), i = 1, \ldots, n-1$, with $v_1 = v$ and $v_n = w$. The weight of the path $W(\pi(v, w))$ is the product of weights of edges from this path:

$$W(\pi(v, w)) := \prod_i c_{v_i, v_{i+1}} \quad (4)$$

Let $\Pi(v, w)$ be the set of all paths connecting nodes $v$ and $w$. And let

$$W(v, w) := \sum_{\pi \in \Pi(v, w)} W(\pi(v, w)) \quad (5)$$

The graph coupled with expressions we consider as a model of regulatory signaling paths system. The distribution of expressions within the system is determined by the topology of the graph, weights and the distribution of expressions of grandfathers.

For example, let $o$ be the grandfather vertex and $M_o^{(P)}$ and $M_o^{(Q)}$ its expressions in these two different states. Denote by $d^2, d$ the variance and standard deviation for grandfather expression in two states, and suppose that they do not depend on the state: $d^2 := \mathrm{Var}(M_o^{(P)}) = \mathrm{Var}(M_o^{(Q)})$. Denote the mean changes in expression of grandfather's gene as $\Delta_o = \mathbb{E} M_o^{(P)} - \mathbb{E} M_o^{(Q)}$. Expression for any non-grandfather vertex $v$ can be expressed for any state $S \in \{P, Q\}$ by the formula:

$$M_v^{(S)} = M_o^{(S)} W(o, v) + \sum_{w \in V \setminus o} \varepsilon_w^{(S)} W(w, v) \quad (6)$$

The mean change in the expression of a gene $v \in V \setminus o$ is given by:

$$\Delta_v := \mathbb{E} M_v^{(P)} - \mathbb{E} M_v^{(Q)} = \Delta_o W(o, v). \quad (7)$$

Moreover, for any $S \in \{P, Q\}$:

$$cov\left(M_v^{(S)}, M_w^{(S)}\right) = d^2 W(o,v) W(o,w) + \sigma^2 \sum_{v' \in V \setminus o} W(v',v) W(v',w) \qquad (8)$$

*Definition.* We say that a pair of genes $v, w \in V$ satisfy **expected correlation inequality** if and only if

$$\Delta_v \Delta_w \, cov\left(M_v^{(P)}, M_w^{(P)}\right) \geq 0, \quad \Delta_v \Delta_w \, cov\left(M_v^{(Q)}, M_w^{(Q)}\right) \geq 0 \qquad (9)$$

If (9) holds then we say that the two gene expressions $M_v^{(P)}, M_w^{(P)}$ or $M_v^{(Q)}, M_w^{(Q)}$ have **expected correlations**. If one or both expected correlations inequalities are not satisfied, we say that $M_v^{(P)}, M_w^{(P)}$ or $M_v^{(Q)}, M_w^{(Q)}$ have **unexpected correlations**.

Note that in the considered model, by (8) the co-variations in (9) do not depend on a state: $cov\left(M_v^{(P)}, M_w^{(P)}\right) = cov\left(M_v^{(Q)}, M_w^{(Q)}\right)$. This independence means that we can use co-variation only in one state in our definition. In this case the following statement takes place.

*Lemma 1.* For any finite DAG network with linear relations between variables there exists some $\sigma_0^2$ such that for any $\sigma^2 < \sigma_0^2$ there are no unexpected correlations into the network.

*Proof.* Direct from formulas (7), (8). By definition (9) and by representations (7), (8) we have:

$$\Delta_v \Delta_w \, cov\left(M_v^{(P)}, M_w^{(P)}\right) =$$
$$\Delta_o^2 W(o,v) W(o,w) \left(d^2 W(o,v) W(o,w) + \sigma^2 \sum_{v' \neq o} W(v',v) W(v',w)\right) =$$
$$\Delta_o^2 d^2 W^2(o,v) W^2(o,w) + \Delta_o^2 \sigma^2 W(o,v) W(o,w) \sum_{v' \neq o} W(v',v) W(v',w) \qquad (10)$$

Here the first term is necessarily positive and the second can be made arbitrarily small by choice of $\sigma^2$. Thus $\Delta_v \Delta_w \, cov\left(M_v^{(P)}, M_w^{(P)}\right)$ can always be made positive (implying that there are no unexpected correlations) by a choice of a sufficiently small variance $\sigma^2$. This statement is precisely Lemma 1.

The formula (8) shows that any link/correlation between two nodes in a network can be represented as a sum of two parts: *causal propagation* from causal node and *noise propagation* part:

$$Cov\left(M_v^{(S)}, M_w^{(S)}\right) = \underbrace{d^2 W(o,v) W(o,w)}_{causal\ propagation} + \underbrace{\sigma^2 \sum_{v' \neq o} W(v',v) W(v',w)}_{noise\ propagation} \qquad (11)$$

Here, it is easy to see that if the grandfather variance $d^2$ increases, then the causal propagation will determine the sign of the covariance after some threshold. It means that it determines a link to be expected or unexpected.

Moreover, Lemma 1 says that if we observe in such regulation networks (DAGs with linear relationships between variables) unexpected correlations, it means that they appeared as a result of noise propagation within the network. Thus, the proportion of unexpected correlation reflects the noise level in a network (to the extent to which this mathematical framework, or that generalized in Section II.2 below, accurately reflects the system being modeled).

*Note 1. The concept of expected correlations was also observed in VanderWeele and Robins, 2010, as a rule governing the relationship between monotonic links and the sign of covariance between variables.*

*Note 2. The linear relations between variables can be generalized: the expression $X_v = f_v(\{X_{v'}\}_{v' \in pa(v)}; \varepsilon_v)$, where $f_v$ is a monotone function, and $\varepsilon_v$ is internal network noise. If structural functions are monotonic function, then the lemma holds also.*

*Estimation of noise.* Error estimation is based on the following: if two genes belong to two independent subnetworks (see Figure 2a), then the correlation between their respective expression levels has to be equal to 0. Observable correlations, however, can be significantly different from 0 due to noise, in which case the observable correlation is positive (or negative) in roughly 50% of the cases (see formula (22)). On average, then, half of all random correlations between any pair of genes from unrelated subnetworks can be classified as unexpected, as in (9). Thus $2 \cdot PUC$ can be used as an estimate of total error.

Moreover, it is possible to prove for tree like graphs that within one network the noise propagation (see the formula (12)) has the same property as stated in formula (22). Indeed, the representation (6) means that any variable $M_v^{(S)}$ can be decomposed into the causal component $X_o^{(S)} W(o, v)$ and the noise component $\xi_v^{(S)} := \sum_{w \in V \setminus o} \varepsilon_w^{(S)} W(w, v)$. Then the covariance between $\xi_v^{(S)}$ and $\xi_w^{(S)}$ can be calculated exactly (compare with formula (10))

$$cov\left(\xi_v^{(S)}, \xi_w^{(S)}\right) = \sigma^2 \sum_{u \in V} W(u, v) W(u, w). \tag{12}$$

If $c_{vw}$ are mutually independent, identically distributed, with positive probabilities for being positive or negative, then the covariance (12) for any $S \in \{P, Q\}$ will be negative approximately in half of all cases.

## II.2. Definitions and generalization.

Here we study the concept of unexpected links in a more general framework. The positive and negative correlation inequalities are an active research direction in the field of probability and statistical mechanics. We believe these inequalities will allow us to generalize the concept of unexpected correlations in the PUC method. The following framework connects FKG (Fortuin–Kasteleyn–Ginibre) inequality in Statistical Mechanics to the concept of expected and unexpected links.

Let $\Omega$ be the underlying sample space of a biological system. As an example of a biological system we consider a gene regulatory network, where $\Omega$ represents the set of all possible gene expression configurations. We can suppose that the state space $\Omega$ has an ordering (or partial ordering) "$\prec$" assigned to pairs of its elements.

*Definition*. *A random variable $X = X(\omega)$ is said to be increasing if $\omega \prec \omega'$ implies $X(\omega) < X(\omega')$. Similarly, a random variable is decreasing if $\omega \prec \omega'$ implies $X(\omega) > X(\omega')$. Both types of random variables, increasing and decreasing, are said to be monotone random variables.*

In the field of statistical mechanics and probabilistic combinatory, the FKG inequality (Fortuin–Kasteleyn–Ginibre inequality) explains most of the results involving monotone random variables and monotone (increasing or decreasing) events. It states that for two increasing random variables $X$ and $Y$,

$$\mathbb{E}(XY) \geq \mathbb{E}(X)\mathbb{E}(Y) \tag{13}$$

In some applications, such as percolation models, partial ordering of $\Omega$ is sufficient for the FKG to hold (Grimmett, 1999). Many important results in applied mathematics and physics, such as the exact value of critical probability in two-dimensional percolation models, would have been impossible without the FKG inequality.

Let $G = (V, E)$ be a graph (network) with vertices (nodes) $V$ and edges $E$. Nodes $v \in V$ represent the genes. Let $X_v(\omega)$ be monotone functions (random variables) assigned to each node $v \in V$. Here $X_v$ represents the noiseless gene expressions. In this framework it is convenient represent the state system as a probability measure. Consider two probability measures $P$ and $Q$ over $\Omega$ such that for all $\omega \in \Omega$:

$$P(\sigma \in \Omega: \sigma \prec \omega) \geq Q(\sigma \in \Omega: \sigma \prec \omega) \tag{14}$$

Here $P$ and $Q$ correspond to the two states of a biological system. Let us denote, as before, $\Delta_v := \mathbb{E}_P[X_v] - \mathbb{E}_Q[X_v]$. We repeat the definition of expected and unexpected links.

*Definition.* *We say that random variables $X_v$ and $X_u$ modeling gene expressions in a pair of genes satisfy* **expected correlation inequality** *if and only if*

$$\Delta_v \Delta_u \, cov_P(X_v, X_u) \geq 0, \quad \Delta_v \Delta_u \, cov_Q(X_v, X_u) \geq 0, \tag{15}$$

*in which case we say that the two gene expressions $X_v$ and $X_u$ have* **expected correlations**. *If one or both expected correlations inequalities are not satisfied, we say that $X_v$ and $X_u$ have* **unexpected correlations**.

*Lemma 2.* *If $X_v$ and $X_u$ are monotone functions, and probability measures $P$ and $Q$ satisfy the condition (13), then $X_v$ and $X_u$ satisfy expected correlation inequality (or $X_v$ and $X_u$ have expected correlations).*

*Proof.* Indeed, if $X_v$ is an increasing (decreasing) variable, then $\Delta_v \leq 0$ ($\Delta_v \geq 0$). Now, if both $X_u$ and $X_v$ are either increasing or decreasing the FKG inequality (13) implies non-negative correlations, so that for any state $S \in \{P, Q\}$

$$cov_S(X_u, X_v) := \mathbb{E}_S[X_u X_v] - \mathbb{E}_S[X_u]\mathbb{E}_S[X_v] \geq 0, \quad \forall u, v \in V, \tag{16}$$

which implies expected correlation inequalities (15).

Similarly, if one of the two variables (i.e. $X_u$ or $X_v$) is increasing while the other is decreasing, the FKG inequality (13) implies non-positive correlations, such that for any state $S \in \{P, Q\}$,

$$cov_S(X_u, X_v) := \mathbb{E}_S[X_u X_v] - \mathbb{E}_S[X_u]\mathbb{E}_S[X_v] \leq 0, \quad \forall u, v \in V \tag{17}$$

implying (15) hold once again. It proves the Lemma 2. □

Next, let $\xi_v$ denote the errors for each node $v \in V$. We assume that the random variables $\xi_v, v \in V$ are functions over a probability space $\Xi$, independent from any probability measure over $\Omega$, such as $P$ and $Q$. Let $\mu$ be the joint distribution of $\xi_v, v \in V$ and $\mathbb{E}_\mu[\xi_v] = 0$ for any $v \in V$. The measured gene expression we quantify as a random variable

$$M_v = X_v + \xi_v, \quad v \in V, \tag{18}$$

over the product space $\Omega \times \Xi$, and the two different states of a biological system correspond to two different probability product measures, $P \times \mu$ and $Q \times \mu$. Note that for any gene $v$:

$$\mathbb{E}_{P \times \mu}[M_v] - \mathbb{E}_{Q \times \mu}[M_v] = \mathbb{E}_P[X_v] - \mathbb{E}_Q[X_v] =: \Delta_v \qquad (19)$$

The following Lemma is an analogous of the Lemma 1 for the general framework.

*Lemma 3.* If the variances of errors $\sigma_v^2 = Var(\xi_v)$ are small enough for all $v \in V$, then the pairs of measured gene expression $M_v$ will also satisfy the inequalities (15). Thus in the noiseless networks we foresee no unexpected correlations.

*Proof.* The proof is a direct consequence of the covariance calculation.

$$cov_{S \times \mu}(M_u, M_v) = cov_{S \times \mu}(X_u + \xi_u, X_v + \xi_v) = cov_S(X_u, X_v) + cov_\mu(\xi_u, \xi_v) \qquad (20)$$

By Cauchy-Schwarz inequality

$$|cov_S(\xi_u, \xi_v)| \le \sigma_u \sigma_v \qquad (21)$$

the second covariance in (19) can be made so small that the sign of $cov_S(M_u, M_v)$ and the sign of $cov_S(X_u, X_v)$ will coincide. This proves Lemma. $\square$

However in the noisy networks, the expected correlations rule (14) can be violated. Here the fraction of edges $(u, v)$ violating (15) that we call the Proportion of the Unexpected Correlations (PUC) becomes an estimator of the frequency of false edges.

### II.3. PUC represents 50% of erroneous.

For any $u, v \in V$; $S \in \{P, Q\}$; and $\mu \in \Xi$, let us assume that the variables $\xi_v$ are random such that, asymptotically, $cov_\mu(\xi_u, \xi_v)$ is positive for half of the $\binom{|V|}{2}$ edges $(u, v)$, and negative for the rest of the pairs:

$$\lim_{|V| \to \infty} \frac{\#\{(u,v): cov_\mu(\xi_u, \xi_v) > 0\}}{\binom{|V|}{2}} = \lim_{|V| \to \infty} \frac{\#\{(u,v): cov_\mu(\xi_u, \xi_v) < 0\}}{\binom{|V|}{2}} = \frac{1}{2}. \qquad (22)$$

If the covariance $cov_{S \times \mu}(M_u, M_v)$ is of a different sign than $cov_S(X_u, X_v)$ (i.e. if a particular correlation $(u, v)$ is unexpected), it must hold that (see (20)):

$$\frac{cov_{S \times \mu}(M_u, M_v) \, cov_S(X_u, X_v)}{\left(cov_\mu(\xi_u, \xi_v)\right)^2} = \left(\frac{cov_S(X_u, X_v)}{cov_\mu(\xi_u, \xi_v)}\right)^2 + \frac{cov_S(X_u, X_v)}{cov_\mu(\xi_u, \xi_v)} < 0. \qquad (23)$$

This condition is of the form $R^2 + R < 0$, where $R = \frac{cov_S(X_u, X_v)}{cov_\mu(\xi_u, \xi_v)}$, which trivially has the solution:

$$\frac{1}{R} = \frac{cov_\mu(\xi_u, \xi_v)}{cov_S(X_u, X_v)} < -1. \tag{24}$$

The resulting inequality is satisfied under two conditions, which are thus requisite for a correlation to be unexpected, namely:

$$|cov_\mu(\xi_u, \xi_v)| > |cov_S(X_u, X_v)| \tag{25}$$

$$cov_\mu(\xi_u, \xi_v) cov_S(X_u, X_v) < 0 \tag{26}$$

The first condition (25) is interpreted as a drowning out of the causal link between two nodes by error; that is, the magnitude of error in the correlation between two nodes' expressions is greater than the magnitude of real correlation between them. The second condition (26) is interpreted as a counteracting of error to causal connections: the contribution to the empirical correlation between two nodes due to error must counteract the contribution due to causal mechanisms.

Condition (26) implies that, given the condition (22) for error distribution, PUC will statistically detect 50% of total false correlations for which the causal contribution is negligibly small, as the signs of the error and causal contribution are equally likely to be the same as they are to be opposite.

### II.4. Unexpected correlations under non-monotonicity.

Here we prove the proposition in the conclusion about non-monotonic links. The statement says that a non-monotonic link between two nodes with an unexpected correlation cannot cause a transition between two distinct states of a network. We provide an extreme example of non-monotonicity, in which the dependence between two nodes changes in sign in the two states of a network (e.g. stimulation in one state of a biological system and inhibition in the other).

Assume we are given $n + 2$ gene expressions in two biological state $P$ and $Q$: $X_P, Y_P, X_{1,P}, \ldots, X_{n,P}$ and $X_Q, Y_Q, X_{1,Q}, \ldots, X_{n,Q}$. We assume linear (or almost linear) dependence of $Y$ on $X$ within any one given biological state, stated as follows: $Y_P = \alpha_P X_P + \xi_P$ and $Y_Q = \alpha_Q X_Q + \xi_Q$, where $\xi_P$ is a function of $X_{1,P}, \ldots, X_{n,P}$, and $\xi_Q$ is a function of $X_{1,Q}, \ldots, X_{n,Q}$, and $\alpha_P \alpha_Q \neq 0$. We suppose that $X_P$ ($X_Q$) and $\xi_P$ ($\xi_Q$) are independent. Recall that all gene expression values are positive and remember that $\Delta X \coloneqq \mathbb{E}_P[X] - \mathbb{E}_Q[X] = \mathbb{E}[X_P] - \mathbb{E}[X_Q]$.

<u>Lemma 4</u>. Suppose $\alpha_P \alpha_Q < 0$ (implying that the relation between X and Y is non-monotonic), then:

  (a) X and Y have unexpected correlations.
  (b) The sign of $\Delta Y$ may not depend on the sign of $\Delta X$, but instead mostly depends on the sign of $\Delta \xi$.

<u>Proof</u>. Observe that, due to independence of $X_P$ ($X_Q$) and $\xi_P$ ($\xi_Q$):

$$cov_P(X,Y) = cov(X_P, Y_P) = \alpha_P Var[X_P], \qquad (27)$$

$$cov_Q(X,Y) = cov(X_Q, Y_Q) = \alpha_Q Var[X_Q]. \qquad (28)$$

Therefore, $cov_P(X,Y) cov_Q(X,Y) < 0$ (so that the expected correlation inequalities do not hold simultaneously) if and only if $\alpha_P \alpha_Q < 0$. This proves the item (a) of the lemma.

Let us prove (b). Without loss of generality, $cov_P(X,Y) < 0$, implying $\alpha_P < 0$ and $\alpha_Q > 0$. Hence:

$$\Delta Y = \mathbb{E}(Y_P - Y_Q) = \mathbb{E}(\alpha_P X_P - \alpha_Q X_Q) + \mathbb{E}(\xi_P - \xi_Q) \qquad (29)$$

Note that $\mathbb{E}(\alpha_P X_P - \alpha_Q X_Q) < 0$ regardless of the values of $X_P$ and $X_Q$ (both of which are strictly positive). Thus in the case $\Delta \xi > 0$ the change $\Delta Y$ will still be negative. The sign of $\Delta Y$ will be positive only if $\Delta \xi \gg 0$.

## III. Simulations using GeneNetWeaver.

We tested PUC using GeneNetWeaver (GNW), a software package designed for rigorous testing of gene network inference methods. We used GNW to generate various networks ranging in size from 40 to 740 nodes, each broken into two disjoint subnetworks in a similar manner as with the previous simulations. Distinct equilibrium network states were made by performing a 50% knockdown on the node in each subnetwork with the most connections. Networks were simulated 100 times both stochastically and analytically. In the case of analytic simulations, in order to get distinct equilibria in different simulations all genes were given normally distributed microperturbations, i.e. proportional up/down regulations with mean 0 and a standard deviation of 1.25%. After each simulation, we selected those genes which were differentially expressed with FDR < 0.01%, and calculated correlations between them in each class separately. We computed PUC and true error for the resulting regulatory networks consisting of at least 20 nodes at various FDR cutoffs. The results are summarized in figures 3S a,b

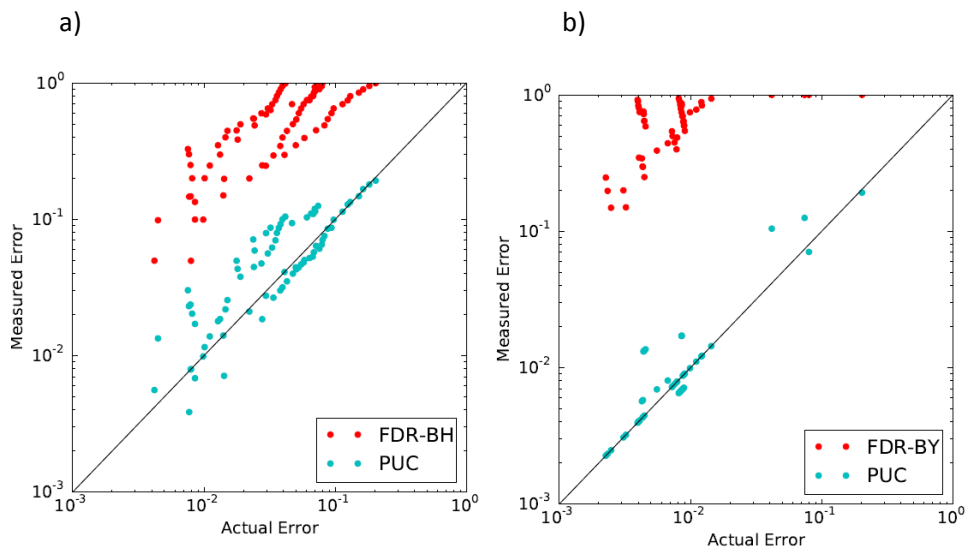

Figure 3S. Comparison between PUC and FDR in networks simulated by GNW.

. x axes represent actual error; y axes actual error (black line), PUC- blue dots, FDR –red dots (Benjamini-Hochberg- left panel; Benjamini-Yekutieli- right panel).

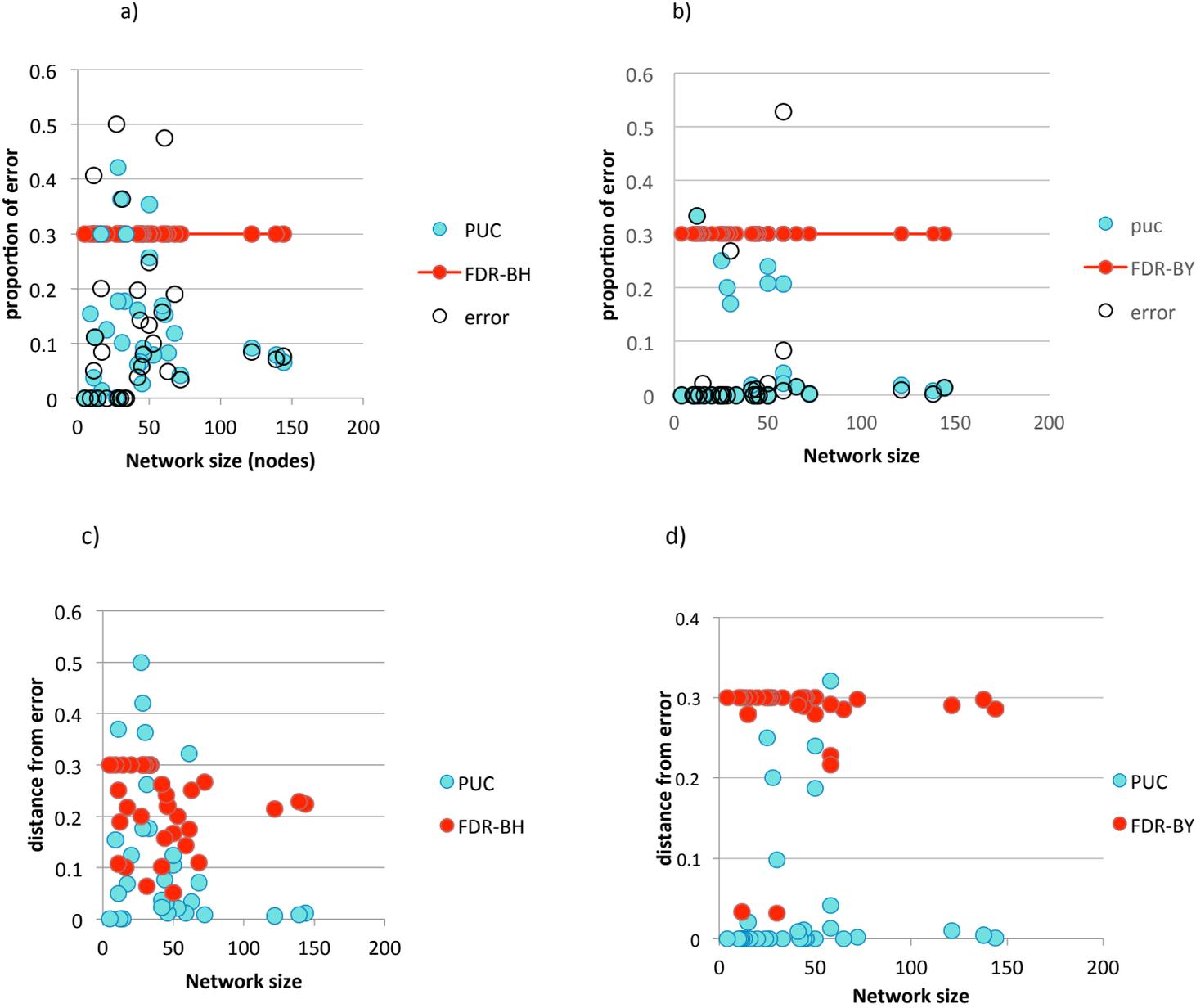

Figure 4S. PUC, FDR-BH (a,c), FDR-BY (b,d) and error in networks of different sizes (number of nodes) simulated by GNW. Panels a) and b) show values for each metric (PUC, FDR or error). Panels c) and d) show the distance from error for FDR and PUC. Overall, PUC is closer to error than FDR.

**List of datasets from BRB Array Tools Archive used for analysis:**

GEO IDs:

GDS1021, GDS232, GDS408, GDS470, GDS484, GDS507, GDS531, GDS535, GDS536, GDS619, GDS690, GDS715, GDS760, GDS806, GDS838, GDS845-8, GDS884, GDS971, GDS978.

Note: for datasets that we could not find GEO ID we provide PUBMED IDs. All datasets were downloaded from BRB Array Tools Archive.

PUBMED IDs:

PMID:10359783, PMID:11707567, PMID:12925757, PMID:15548776, PMID:11707590.